\documentclass[aps,prl,twocolumn,groupedaddress,showpacs,floatfix,altaffilletter]{revtex4-1}
\usepackage{graphicx}
\usepackage{grffile}
\usepackage{amsmath}
\usepackage{amssymb}
\usepackage{bm}
\usepackage{color}
\usepackage[dvipsnames]{xcolor}
\usepackage{amsmath,amssymb,amsfonts}
\usepackage{epsfig}
\usepackage{times}
\usepackage[colorlinks,bookmarks=false,citecolor=blue,linkcolor=red,urlcolor=blue]{hyperref}

\newcommand{\fref}[1]{Fig.~\ref{#1}}
\newcommand{\eref}[1]{Eq.~(\ref{#1})}

\newcommand{\normwidth}{0.8\columnwidth}

\newcommand{\figwidth}{0.97\columnwidth}

\newcommand{\cred}{\color{black}}

\begin{document}

\title{Electric polarization of Sr$_{0.5}$Ba$_{0.5}$MnO$_{3}$: a multiferroic Mott insulator }

\author{R.~Nourafkan$^{1}$, G.~Kotliar$^{2}$, and A.-M.S. Tremblay$^{1,3}$}
\affiliation{$^1$D{\'e}partement de Physique and RQMP, Universit{\'e} de Sherbrooke, Sherbrooke, Qu{\'e}bec, Canada J1K 2R1}
\affiliation{$^{2}$Department of Physics \& Astronomy, Rutgers University, Piscataway, NJ 08854-8019, USA}
\affiliation{$^3$Canadian Institute for Advanced Research, Toronto, Ontario, Canada}
\date{\today}
\begin{abstract}

Ab initio calculations of the electric polarization of correlation-driven insulating materials, namely Mott insulators, have not been possible so far. 
Using a combination of density functional theory and dynamical-mean-field theory we study the electric polarization of the Mott insulator Sr$_{0.5}$Ba$_{0.5}$MnO$_{3}$. We predict a ferroelectric polarization of $\simeq 16.5~\mu C/cm^2$ in the high temperature paramagnetic phase and recover the measured value of $\simeq 13.3~\mu C/cm^2$ in the low temperature antiferromagnetic phase. 
Our calculations reveal that the driving force for the ferroelectricity, the hybridization between Mn e$_g$ and O p orbitals, is suppressed by correlations, in particular by the Hund coupling and by the onset of magnetic order. 
They also confirm that the half-filled Mn $t_{2g}$ orbitals give rise to the antiferromagnetic Mott phase. 
This magnetic ordering leads to changes in the ionic polar displacement and in turn to the electronic polarization. In addition, for a fixed ionic displacement, we find that there is a reduction in the electronic contribution due to partial magnetic polarization of the Mn e$_g$ orbitals. The reduction of the polarization due to ionic displacement
dominates over the additional electronic part, hence the net magneto-electric coupling is negative.

\end{abstract}
%77.80.-e 	Ferroelectricity and antiferroelectricity
%71.10.-w 	Theories and models of many-electron systems
%71.27.+a 	Strongly correlated electron systems; heavy fermions
%75.47.Lx 	Magnetic oxides

\maketitle

Multiferroics, materials which display simultaneous magnetic and ferroelectric orders,~\cite{Khomskii20061} are interesting both for their rich physics and for their promising practical applications. The search for materials with strong-magneto-electric coupling is challenging and requires an understanding of how the magnetic order, or more specifically the correlations, influence the electric polarization and vice versa.

Hexagonal multiferroics like $R$MnO$_3$, or Bi- or Pb-based ferroelectrics, like BiFeO$_3$, all have strong e-e correlations on the Mn or Fe site. Due to the independent origin of ferroelectricity and magnetism in these types of multiferroics,~\cite{PhysRevLett.98.217601, PhysRevB.71.014113} it has been argued that while the electronic correlations govern the magnetic order, they might not have a profound effect on the electric polarization. However, this is not a completely satisfactory argument and further investigation is necessary to understand the relation between magnetism and polarization and, in particular, the impact of correlations on these quantities. Indeed, recent observations on magnetoelectric effects in BiFeO$_3$ suggest a sizable Fe electric polarization which is suppressed by the onset of the magnetic transition.~\cite{PhysRevB.88.060103} 

From a theoretical point of view, calculations of the electric polarization in the paramagnetic (PM) insulating phase of multiferroics is essential to answer a number of questions, including, as mentioned above, that of the mutual influence of magnetic order and electric polarization. However, the electric polarization of multiferroic materials in their PM insulating phase has not been thoroughly investigated so far, partly due to the absence of a theory to calculate this quantity and partly due to the difficulty of measuring it. 

Theoretical studies of the electric polarization are based on the modern theory of polarization, also known as the Berry phase approach, which is a single-electron theory.~\cite{PhysRevB.47.1651} While spin-polarized band-structure calculations with interactions treated in the Hartree-Fock approximation  may be used to partially take into account correlation effects in the antiferromagnetic (AF) insulating phase,~\cite{PhysRevLett.100.227603, PhysRevB.71.014113, PhysRevB.63.045111} the correlation-driven insulating PM state is beyond the scope of this approach. Indeed, the PM state in these approaches is a metal, and hence it cannot be ferroelectric. Here we show that combining correlated band structure calculations (DFT+DMFT) with a recently proposed formula for the electric polarization of interacting insulators, expressed in terms of the full Green and vertex functions,~\cite{PhysRevB.88.155121} provides a practical and reliable way to calculate the polarization in the PM phase. This opens the way towards a better understanding of correlated multiferroic materials and, eventually, towards engineering materials with large magnetoelectric responses.

Here we focus on the Mott insulator Sr$_{1-x}$Ba$_{x}$MnO$_{3}$, where it has recently been experimentally shown that a 
transition to a ferroelectric phase occurs for $x \geq 0.45$.~\cite{PhysRevLett.107.137601, PhysRevB.86.104407}  In this compound both magnetic and ferroelectric instabilities are related to the Mn ions. In the ferroelectric phase the crystal structure changes from cubic to tetragonal: the elongation of the $c$ axis allows for a displacement of magnetic Mn$^{4+}$ ions from the center of the surrounding oxygen octahedron leading to a Mn-O-Mn bond angle deviation far from ideal $180^{\circ}$. The ferroelectric transition temperature $T_C$ is $~410~K$ for $x=0.5$, the stoichiometric case we consider in this paper. Below $T_N(<T_C)$, the system shows a multiferroic phase associated with G-type AF ordering of off-center Mn$^{4+}$ with d$^3$ configuration. 
The onset of the low-temperature long-range AF ordering reduces the Mn-O-Mn bond deviation from $180^{\circ}$ and suppresses the off-centering displacement of the Mn$^{4+}$ ion.~\cite{PhysRevLett.107.137601} 
Microscopically, the magnetic superexchange interactions between Mn $t_{2g}$ orbitals tends to suppress the Mn-O-Mn bond angle deviation and hence reduces the ionic ferroelectric displacements.~\cite{PhysRevLett.109.107601} This reduction decreases the ionic contribution to the polarization, therefore resulting in a negative magnetoelectric coupling.~\cite{PhysRevLett.109.107601}

Here we address the impact of correlations and magnetic order on the ferroelectricity of the system by investigating the electronic contribution to the electric polarization in both the PM and AF phases.  

\paragraph{Method} 

The change of electric polarization of a material in response to an inversion-symmetry-breaking distortion has two contributions, one purely ionic and one from the electronic responses. The ionic part is $\Delta {\bf P}^{\rm ion}=(e/V_{\rm cell})\sum_{i=1}^{N_{\rm atom}}Z_i^{\rm ion}\cdot({\bf r}_i-{\bf r}_i^{0})$ where $V_{{\rm cell}}$ is the cell volume, $e$ is elementary charge, $Z_i^{\rm ion}$ is the ionic charge, equal to the number of valence electrons in the atom $i$, and ${\bf r}_i$ (${\bf r}^{0}_i$)  is the position vector for the polar (centro-symmetric) structure. The electronic part can be obtained by integrating the bulk transient current as the system adiabatically evolves from a centro-symmetric structure to the polar structure. In the presence of the electronic correlations it is given by ~\cite{PhysRevB.88.155121} 
%
%\begin{widetext}
\begin{align}
\Delta {\bf P}^{\rm el}_i&=\int_0^1 d\xi (\frac{-ie}{6NV_{{\rm cell}}})\sum_{{\bf k}}(\frac{1}{\beta})\sum_{\omega_m}
\varepsilon^{k_0k_1k_2}\nonumber\\&
{\rm Tr}\bigg({\bf G} \partial_{k_0}{\bf G}^{-1}
{\bf G} \partial_{k_1}{\bf G}^{-1}
{\bf G} \partial_{k_2}{\bf G}^{-1}
\bigg),\label{eq:Pol}
\end{align}
%\end{widetext}
%
where $\beta$ is the inverse temperature, $k_0,k_1,k_2 \in \{k_i,i\omega_m,\xi \}$ and $\xi$ is a dimensionless parameter that parametrizes an inversion-symmetry-breaking distortion of the system. The interacting single-particle Green function entering \eref{eq:Pol} is
\begin{equation}
{\bf G}({\bf k},i\omega_m)=[(i\omega_m+\mu){\bm 1}-{\bf H}_0({\bf k})-{\bm \Sigma}({\bf k},i\omega_m)]^{-1}, 
\end{equation}
where ${\bf H}_0$ denotes the non-interacting part of Hamiltonian, ${\bm \Sigma}$ is the electron self-energy, $\mu$ is the chemical potential and $\omega_m$ denotes the Matsubara frequencies. 
Bold quantities are $n\times n$ matrices where $n$ denotes the number of orbitals within the unit cell. In \eref{eq:Pol}, vertices, $\partial_{k_0}{\bf G}^{-1}$, have two contributions: the bare vertex, describing the response of a noninteracting system, and the vertex corrections, arising from the many-body physics contained in the self-energy.   

Although the trace in \eref{eq:Pol} can be evaluated in an arbitrary basis, it is more convenient to work in a distortion-independent orbital basis. \footnote{It is easier to obtain the distortion vertex, $\partial_{\xi}{\bf G}^{-1}$, when the Green function matrix is written in a basis which is independent of $\xi$ along the path connecting the centro-symmetric system ($\xi=0$) to the real system ($\xi=1$).} In this basis, the non-interacting part of the Hamiltonian is given by~\cite{PhysRevB.80.085101} 

\begin{equation}
[{\bf H}_0^{\sigma}({\bf k})]_{\alpha_1 \gamma_1,\alpha_2 \gamma_2} = \sum_{n n'}{\bf P}_{\alpha_1 \gamma_1,n}^{\sigma}({\bf k})\epsilon^{\sigma}_{n}({\bf k})\delta_{n n'}[{\bf P}_{\alpha_2  \gamma_2,n'}^{\sigma}({\bf k})]^*,
\end{equation}
where $\epsilon^{\sigma}_{n}({\bf k})$'s are the eigenvalues of the LDA Hamiltonian and 
${\bf P}_{\alpha_1 \gamma_1,n}^{\sigma}({\bf k})\equiv \langle \chi_{\alpha_1 \gamma_1}^{\sigma}|\psi^{\sigma}_{{\bf k}n}\rangle$
are the matrix elements of the projectors from the Kohn-Sham basis set, $|\psi^{\sigma}_{{\bf k}n}\rangle$, to the atomic-like Wannier basis $|\chi_{\alpha_1 \gamma_1}^{\sigma} \rangle$. $\alpha$ is the index of the atom within the unit cell, $\gamma$ is a short-hand orbital index, and $\sigma$ is the spin degree of freedom. This basis is constructed from the Kohn-Sham bands contained within a suitable energy window for the $d$ shell of Mn, and $p$-shell of O.\footnote{We emphasize that we have an orthonormal projection. In either AF or PM phases, the number of orbitals is equal to the number of bands within the energy window.} The bare current vertices can be obtained by employing the definition of the dipole matrix elements versus momentum operator.~\cite{PhysRevB.85.205133} 
The same prescription as above is used to write the current vertices in the orbital basis.

We use the theoretical framework that combines dynamical mean-field theory (DMFT)~\cite{RevModPhys.68.13} with a continuous-time quantum Monte-Carlo impurity solver~\cite{PhysRevB.74.155107, PhysRevB.75.155113} and
density-functional theory in the local density approximation (LDA), in the charge self-consistent implementation based on the WIEN2K package.~\cite{PhysRevB.81.195107} For the $3d$ states of Mn, we use a Hubbard interaction $U=5.0~eV$ and a Hund's rule coupling $J=0.7~eV$, unless stated otherwise.
Calculations in the AF phase are done at $T=100~K$, while in PM phase they are done at $T=225~K$. We also perform calculations within the LSDA+U scheme, obtaining the ferroelectric polarization from the Berry phase approach.

\paragraph{Electronic structures of PM and AF phase}

The crystal structures are fixed to the experimental structures for Sr$_{0.5}$Ba$_{0.5}$MnO$_3$.~\cite{PhysRevLett.107.137601} The $c/a$ ratio, Mn-apical O dimerization, and the Mn-planar O-Mn bond angle are, respectively, $~1.003$, $~0.009c$ and $179.1^{\circ}$ in the AF phase, while in the PM phase they are $~1.012$, $~0.018c$ and $175.4^{\circ}$. 

\begin{figure}
%\begin{center}
\includegraphics[width=\figwidth]{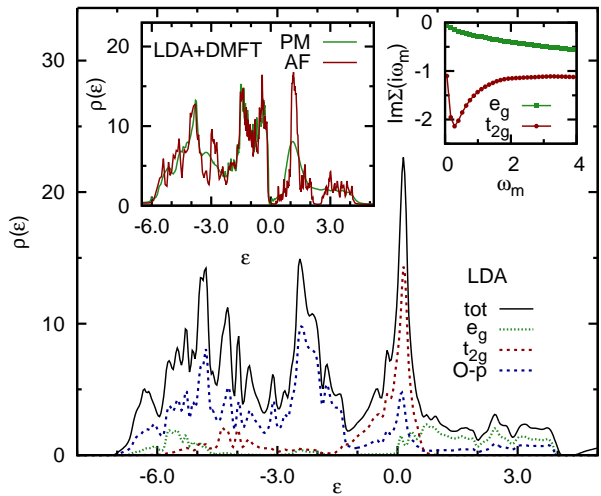}
\caption{(Color online) The total and orbitally resolved LDA paramagnetic DOS. Insets: Predicted LDA+DMFT(PM) DOS and spin-resolved LDA+DMFT(AF) DOS and imaginary part of the electron self-energy as a function of Matsubara frequency. The analytical continuation for the LDA+DMFT method has been performed with the Maximum Entropy method as implemented in Ref.\onlinecite{PhysRevB.81.195107}. }\label{fig:LDA_dos}
%\end{center}
\end{figure}

The single-particle density of states (DOS), the site-projected DOS of the Mn $t_{2g}$, Mn $e_{g}$, and O $p$  states, calculated within LDA are shown in the main panel of \fref{fig:LDA_dos} for the high-temperature crystal structure.~\footnote{The distortion in the ferroelectric phase lowers the local symmetry of the lattice, thus lifting the degeneracies of the $t_{2g}$- and $e_g$-orbitals. However, the energy splitting is very small due to the smallness of the distortion and we keep using this terminology.} The LDA-DOS within the energy window $[-7.5,4.5]~eV$  around the Fermi energy arises from the Mn $3d$ states and oxygen $2p$ valence orbitals. The narrow band around the Fermi energy arises predominantly from the Mn $t_{2g}$ states. Lying below and hybridized with the Mn states is the broad predominantly oxygen $2p$ valence band. The Mn $e_g$ states appear predominantly above the Fermi level. The oxygen weight in the $t_{2g}$ and $e_g$ energy window can be used as a measure of covalency or hybridization. \fref{fig:LDA_dos} shows that Mn $e_{g}$ and O $p$ hybridize above the Fermi level, but also at high binding energy in the valence bands. Due to the latter hybridization, the $e_g$ states are partially filled. 

The effect of correlations is illustrated by the  spectral function, obtained from the LDA+DMFT(PM) calculations, plotted in the inset of \fref{fig:LDA_dos}. The main changes compared with LDA occur near the Fermi level. One observes a small correlation-driven insulating gap. The Mn $e_g$ states are pushed to higher energies while the $t_{2g}$ states split into lower and upper Hubbard bands, lying below and above the Fermi level respectively. The oxygen-$p$ bands do not shift appreciably with respect to the LDA calculation. Hence, the lower Hubbard band, located at $\simeq -0.35~eV$, has an admixture of oxygen-$p$. 

The Mott nature of the insulating gap can be verified from the Matsubara frequency dependence of the imaginary part of the self-energy shown in the inset of \fref{fig:LDA_dos}. The features for the $t_{2g}$-orbitals are characteristic of poles in the gap, while the smooth Fermi-liquid like behavior for the $e_g$ self-energy suggests moderate correlation effects. 
The half-filled configuration of $t_{2g}$ shell makes them active to respond to the effects of electronic correlations.

Spin-polarized LDA (LSDA) calculations for the low-temperature crystal structure show that the system is insulating, with a small gap of $\simeq 0.4~eV$. 
The gap is only slightly modified if we add the interactions either through the LSDA+U method,~\footnote{In our LSDA+U calculation, only an effective $U_{eff}=U-J$ enters the Hamiltonian.} or through LDA+DMFT(AF). In the LSDA+U method some spectral weight shifts to higher energies, while in the LDA+DMFT(AF), shown in the inset of \fref{fig:LDA_dos}, the main peak above the Fermi level decreases slightly in energy.

The Mn$^{4+}$ ion has two empty $e_g$-orbitals and three half-filled $t_{2g}$-orbitals. The atomic Hund's rule leads to the high-spin ground state for the Mn$^{4+}$ ions with a formal value of $3.0~\mu_B$ for the magnetic moment. {\cred Due to the finite bandwidth of the $3d$ states, the magnetic moment of Mn is reduced from the formal value to a value of $~2.37~\mu_B$ in LSDA and to slightly larger values of $~2.75~\mu_B$ and $~2.53~\mu_B$ when the localizing effect of $U$ and $J$ are taken into account in the LSDA+U or LDA+DMFT(AF) methods, respectively.} \footnote{The magnetic moment in LSDA+U method is larger than in LDA+DMFT(AF) because of the more localized atomic-like Wannier basis used in LSDA+U, leading to a weaker screening of the interaction.~\cite{PhysRevB.81.195107}} 

\paragraph{ Ferroeletric polarization in the PM and AF phases}

The bare distortion-current vertex, $\partial_{\xi}{\bf H}_0$, provides a measure of the current flowing between different orbitals while the ions are adiabatically displaced. Our calculation shows that in both PM and AF phases the largest matrix element is between Mn $e_g$ and O $p$. We also find that the magnetic and ferroelectric orders are predominately coupled with different orbitals. The driving force behind the ferroelectric instability is the hybridization between the empty Mn $e_g$ orbitals and the O $p$-orbitals~\cite{PhysRevB.83.054110,PhysRevB.79.205119}, while the half-filled Mn $t_{2g}$ orbitals are responsible for the magnetism.~\cite{PhysRevLett.109.107601} The hybridization between empty $e_g$ orbitals and O $p$-orbitals has been recognized in other compounds as a mechanism for the ferroelectric instability.~\cite{arXiv:1404.7705}
The distortion vertex corrections, $\partial_{\xi}{\bm \Sigma}$, take into account the effect of correlations on the distortion-induced current. We find small distortion vertex corrections. 
 
We evaluate the ferroelectric polarization as summation of the ionic and electronic contributions. Our results within LSDA and LSDA+U compare well with other ab-initio calculations on the same compounds ~\cite{PhysRevLett.109.107601,JPhysCondensMatter26215401} and we find the electronic contribution pointing in the direction opposite to the ionic one.
The ionic contribution due to the ferroelectric displacements in the PM phase is larger if compared to the AF phase. 

\begin{figure}
%\begin{center}
\includegraphics[width=\normwidth]{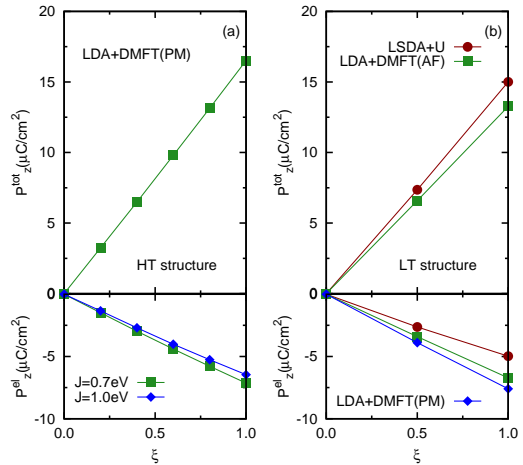}
\caption{(Color online) Total (top) and electronic (bottom) electric polarization in the $z$ direction as a function of inversion symmetry-breaking distortion $\xi$ calculated for the PM phase for two values of $J$ in the high-temperature (HT) structure (panel (a)) and for the AF phase in the low-temperature (LT) structure (panel (b)). Panel (b) also shows the electric polarization in PM phase of the LT structure. $\xi=1$ denotes the experimental structure.  }\label{fig:P_AFPM}
%\end{center}
\end{figure}

The top part of panel (a) of \fref{fig:P_AFPM} shows the total electric polarization in the $z$ direction, $P^{tot}_z$, as a function of the distortion $\xi$ in the PM phase using LDA+DMFT(PM). The electronic contribution is in the lower part. The calculated value in the fully distorted system, $\xi=1$, is $\simeq 16.5~\mu C/cm^2$. 

The off-centering of Mn ions towards O ions create a strong covalent bond of Mn $e_g$-O $p$ which implies the formation of a singlet state opposite to the Hund's rule exchange between localized $t_{2g}$-electrons and the $e_g$-electrons participating in the bond.~\cite{Khomskii20061} 
The role of Hund's coupling in the depletion of the singlet state is confirmed by the results displayed in the left panels of \fref{fig:P_AFPM}, where we see that the electronic polarization is suppressed by a larger $J$ in the PM phase (see the lower panel).   

Panel (b) of \fref{fig:P_AFPM} shows the calculated electric polarization in the $z$ direction, $P^{tot}_z$, in the AF phase using both LSDA+U and LDA+DMFT(AF). All results are in quantitative agreement with the experimental result, $P^{tot}_z=13.5\mu C/cm^2$.~\cite{PhysRevLett.107.137601} The calculated value in LDA+DMFT(AF) is $P^{tot}_z=13.3\mu C/cm^2$. 
Using the same interaction strengths, the LSDA+U method yields a smaller electronic ferroelectricity in comparison to the LDA+DMFT(AF) calculation. This is a consequence of the displacement of the spectral weight
away from the band gap edges. The LDA+DMFT(AF) calculation shows weak
bandwidth renormalization, justifying the relative success of the LSDA+U method.

To further assess the interplay between the ferroelectric and magnetic orderings we calculate the polarization of the low temperature structure (which is normally AF) in the PM phase. This allows us to study the impact of magnetic order on the electronic ferroelectricity using a fixed crystal structure. The results are in the lower part of panel (b) of \fref{fig:P_AFPM}. 
The smaller absolute value of the electronic ferroelectricity in the AF phase is consistent with the singlet breaking role of Hund's coupling~\cite{Khomskii20061,PhysRevLett.109.107601} and decreasing of the hybridization with the $p$ orbitals, which is a measure of the the electronic ferroelectricity. In the AF phase, the Mn $e_g$ orbitals are
partially magnetically polarized, decreasing the hybridization
with the O $p$ orbitals and hence the electronic ferroelectricity as
well. Therefore, the onset of the magnetic order suppresses both ionic and, to a lower extent, electronic ferroelectricity. Note that if the structure did not change between the AF and PM phases, the magneto-electric response would be positive since, as shown in ~\fref{fig:P_AFPM}, the electronic polarization is reduced in the AF phase compared to the PM phase leading to a larger total electric polarization.

\paragraph{Conclusion:} Using LDA+DMFT and a recently proposed formula based on the Green function formalism to calculate the electronic contribution to the polarization of correlated insulators, we investigated the electric polarization and the magneto-electric coupling of the polar Mott insulator Sr$_{0.5}$Ba$_{0.5}$MnO$_{3}$ in both the PM and AF phases. 
In agreement with experiment, we find a ferroelectric polarization of $\simeq 13.3~\mu C/cm^2$ in the AF phase (at $T=100 K$) and we predict a larger ferroelectric polarization of $\simeq 16.5~\mu C/cm^2$ in the PM phase (at $T=225 K$). 
We propose the orbitally-resolved distortion-current vertex as a measure of 
the driving force for ferroelectricity. 
In agreement with the LSDA analysis of Wannier centers \cite{PhysRevB.83.054110} we conclude that the driving force behind the ferroelectric distortion comes from the tendency of Mn $e_g$ states to establish a stronger covalency with the surrounding oxygens. This covalency is reduced by correlations, in particular by Hund coupling. On the other hand, the half-filled Mn $t_{2g}$ orbitals give rise to the magnetic ordering  \cite{PhysRevLett.109.107601} which decreases the ionic displacement, hence its contribution to the polarization. For fixed ionic displacement, the magnetic order also slightly decreases the electronic contribution to the electric polarization by partially polarizing the Mn $e_g$ orbitals through Hund coupling.  Despite this last effect, which would lead to positive magneto-electric coupling, the combination of the two effects gives a negative magneto-electric coupling.
Our calculation of the electric polarization of Sr$_{0.5}$Ba$_{0.5}$MnO$_{3}$ in both the PM and AF phases adds the missing piece to the understanding of the interplay between ferroelectric and magnetic properties of such materials and should serve as a guideline in the search of materials with strong magneto-electric coupling.

\begin{acknowledgments}
R.~N is grateful to J.~Tomczak, and M.~Aichhorn for useful discussions and to K.~Haule for discussion and use of his code. We also acknowledge G. Giovannetti for suggesting the compound, critically reading the manuscript and for his insightful comments. This work has been supported by the Natural Sciences and Engineering Research Council of Canada (NSERC), the Tier I Canada Research Chair Program (A.-M.S.T.) and by NSF Grant No. DMR-1308141. Simulations were performed on computers provided by CFI, MELS, Calcul Qu\'ebec and Compute Canada.

\end{acknowledgments}


\begin{thebibliography}{26}%
\makeatletter
\providecommand \@ifxundefined [1]{%
 \@ifx{#1\undefined}
}%
\providecommand \@ifnum [1]{%
 \ifnum #1\expandafter \@firstoftwo
 \else \expandafter \@secondoftwo
 \fi
}%
\providecommand \@ifx [1]{%
 \ifx #1\expandafter \@firstoftwo
 \else \expandafter \@secondoftwo
 \fi
}%
\providecommand \natexlab [1]{#1}%
\providecommand \enquote  [1]{``#1''}%
\providecommand \bibnamefont  [1]{#1}%
\providecommand \bibfnamefont [1]{#1}%
\providecommand \citenamefont [1]{#1}%
\providecommand \href@noop [0]{\@secondoftwo}%
\providecommand \href [0]{\begingroup \@sanitize@url \@href}%
\providecommand \@href[1]{\@@startlink{#1}\@@href}%
\providecommand \@@href[1]{\endgroup#1\@@endlink}%
\providecommand \@sanitize@url [0]{\catcode `\\12\catcode `\$12\catcode
  `\&12\catcode `\#12\catcode `\^12\catcode `\_12\catcode `\%12\relax}%
\providecommand \@@startlink[1]{}%
\providecommand \@@endlink[0]{}%
\providecommand \url  [0]{\begingroup\@sanitize@url \@url }%
\providecommand \@url [1]{\endgroup\@href {#1}{\urlprefix }}%
\providecommand \urlprefix  [0]{URL }%
\providecommand \Eprint [0]{\href }%
\providecommand \doibase [0]{http://dx.doi.org/}%
\providecommand \selectlanguage [0]{\@gobble}%
\providecommand \bibinfo  [0]{\@secondoftwo}%
\providecommand \bibfield  [0]{\@secondoftwo}%
\providecommand \translation [1]{[#1]}%
\providecommand \BibitemOpen [0]{}%
\providecommand \bibitemStop [0]{}%
\providecommand \bibitemNoStop [0]{.\EOS\space}%
\providecommand \EOS [0]{\spacefactor3000\relax}%
\providecommand \BibitemShut  [1]{\csname bibitem#1\endcsname}%
\let\auto@bib@innerbib\@empty
%</preamble>
\bibitem [{\citenamefont {Khomskii}(2006)}]{Khomskii20061}%
  \BibitemOpen
  \bibfield  {author} {\bibinfo {author} {\bibfnamefont {D.}~\bibnamefont
  {Khomskii}},\ }\href {\doibase http://dx.doi.org/10.1016/j.jmmm.2006.01.238}
  {\bibfield  {journal} {\bibinfo  {journal} {Journal of Magnetism and Magnetic
  Materials}\ }\textbf {\bibinfo {volume} {306}},\ \bibinfo {pages} {1 }
  (\bibinfo {year} {2006})}\BibitemShut {NoStop}%
\bibitem [{\citenamefont {Cho}\ \emph {et~al.}(2007)\citenamefont {Cho},
  \citenamefont {Kim}, \citenamefont {Park}, \citenamefont {Rho}, \citenamefont
  {Park}, \citenamefont {Noh}, \citenamefont {Kim}, \citenamefont {Oh},
  \citenamefont {Park}, \citenamefont {Ahn}, \citenamefont {Ishibashi},
  \citenamefont {Cheong}, \citenamefont {Lee}, \citenamefont {Murugavel},
  \citenamefont {Noh}, \citenamefont {Tanaka},\ and\ \citenamefont
  {Jo}}]{PhysRevLett.98.217601}%
  \BibitemOpen
  \bibfield  {author} {\bibinfo {author} {\bibfnamefont {D.-Y.}\ \bibnamefont
  {Cho}}, \bibinfo {author} {\bibfnamefont {J.-Y.}\ \bibnamefont {Kim}},
  \bibinfo {author} {\bibfnamefont {B.-G.}\ \bibnamefont {Park}}, \bibinfo
  {author} {\bibfnamefont {K.-J.}\ \bibnamefont {Rho}}, \bibinfo {author}
  {\bibfnamefont {J.-H.}\ \bibnamefont {Park}}, \bibinfo {author}
  {\bibfnamefont {H.-J.}\ \bibnamefont {Noh}}, \bibinfo {author} {\bibfnamefont
  {J.}~\bibnamefont {Kim}, \bibfnamefont {B.}}, \bibinfo {author}
  {\bibfnamefont {S.-J.}\ \bibnamefont {Oh}}, \bibinfo {author} {\bibfnamefont
  {H.-M.}\ \bibnamefont {Park}}, \bibinfo {author} {\bibfnamefont {J.-S.}\
  \bibnamefont {Ahn}}, \bibinfo {author} {\bibfnamefont {H.}~\bibnamefont
  {Ishibashi}}, \bibinfo {author} {\bibfnamefont {S.-W.}\ \bibnamefont
  {Cheong}}, \bibinfo {author} {\bibfnamefont {J.~H.}\ \bibnamefont {Lee}},
  \bibinfo {author} {\bibfnamefont {P.}~\bibnamefont {Murugavel}}, \bibinfo
  {author} {\bibfnamefont {T.~W.}\ \bibnamefont {Noh}}, \bibinfo {author}
  {\bibfnamefont {A.}~\bibnamefont {Tanaka}}, \ and\ \bibinfo {author}
  {\bibfnamefont {T.}~\bibnamefont {Jo}},\ }\href {\doibase
  10.1103/PhysRevLett.98.217601} {\bibfield  {journal} {\bibinfo  {journal}
  {Phys. Rev. Lett.}\ }\textbf {\bibinfo {volume} {98}},\ \bibinfo {pages}
  {217601} (\bibinfo {year} {2007})}\BibitemShut {NoStop}%
\bibitem [{\citenamefont {Neaton}\ \emph {et~al.}(2005)\citenamefont {Neaton},
  \citenamefont {Ederer}, \citenamefont {Waghmare}, \citenamefont {Spaldin},\
  and\ \citenamefont {Rabe}}]{PhysRevB.71.014113}%
  \BibitemOpen
  \bibfield  {author} {\bibinfo {author} {\bibfnamefont {J.~B.}\ \bibnamefont
  {Neaton}}, \bibinfo {author} {\bibfnamefont {C.}~\bibnamefont {Ederer}},
  \bibinfo {author} {\bibfnamefont {U.~V.}\ \bibnamefont {Waghmare}}, \bibinfo
  {author} {\bibfnamefont {N.~A.}\ \bibnamefont {Spaldin}}, \ and\ \bibinfo
  {author} {\bibfnamefont {K.~M.}\ \bibnamefont {Rabe}},\ }\href {\doibase
  10.1103/PhysRevB.71.014113} {\bibfield  {journal} {\bibinfo  {journal} {Phys.
  Rev. B}\ }\textbf {\bibinfo {volume} {71}},\ \bibinfo {pages} {014113}
  (\bibinfo {year} {2005})}\BibitemShut {NoStop}%
\bibitem [{\citenamefont {Lee}\ \emph {et~al.}(2013)\citenamefont {Lee},
  \citenamefont {Fernandez-Diaz}, \citenamefont {Kimura}, \citenamefont {Noda},
  \citenamefont {Adroja}, \citenamefont {Lee}, \citenamefont {Park},
  \citenamefont {Kiryukhin}, \citenamefont {Cheong}, \citenamefont {Mostovoy},\
  and\ \citenamefont {Park}}]{PhysRevB.88.060103}%
  \BibitemOpen
  \bibfield  {author} {\bibinfo {author} {\bibfnamefont {S.}~\bibnamefont
  {Lee}}, \bibinfo {author} {\bibfnamefont {M.~T.}\ \bibnamefont
  {Fernandez-Diaz}}, \bibinfo {author} {\bibfnamefont {H.}~\bibnamefont
  {Kimura}}, \bibinfo {author} {\bibfnamefont {Y.}~\bibnamefont {Noda}},
  \bibinfo {author} {\bibfnamefont {D.~T.}\ \bibnamefont {Adroja}}, \bibinfo
  {author} {\bibfnamefont {S.}~\bibnamefont {Lee}}, \bibinfo {author}
  {\bibfnamefont {J.}~\bibnamefont {Park}}, \bibinfo {author} {\bibfnamefont
  {V.}~\bibnamefont {Kiryukhin}}, \bibinfo {author} {\bibfnamefont {S.-W.}\
  \bibnamefont {Cheong}}, \bibinfo {author} {\bibfnamefont {M.}~\bibnamefont
  {Mostovoy}}, \ and\ \bibinfo {author} {\bibfnamefont {J.-G.}\ \bibnamefont
  {Park}},\ }\href {\doibase 10.1103/PhysRevB.88.060103} {\bibfield  {journal}
  {\bibinfo  {journal} {Phys. Rev. B}\ }\textbf {\bibinfo {volume} {88}},\
  \bibinfo {pages} {060103} (\bibinfo {year} {2013})}\BibitemShut {NoStop}%
\bibitem [{\citenamefont {King-Smith}\ and\ \citenamefont
  {Vanderbilt}(1993)}]{PhysRevB.47.1651}%
  \BibitemOpen
  \bibfield  {author} {\bibinfo {author} {\bibfnamefont {R.~D.}\ \bibnamefont
  {King-Smith}}\ and\ \bibinfo {author} {\bibfnamefont {D.}~\bibnamefont
  {Vanderbilt}},\ }\href {\doibase 10.1103/PhysRevB.47.1651} {\bibfield
  {journal} {\bibinfo  {journal} {Phys. Rev. B}\ }\textbf {\bibinfo {volume}
  {47}},\ \bibinfo {pages} {1651} (\bibinfo {year} {1993})}\BibitemShut
  {NoStop}%
\bibitem [{\citenamefont {Giovannetti}\ and\ \citenamefont {van~den
  Brink}(2008)}]{PhysRevLett.100.227603}%
  \BibitemOpen
  \bibfield  {author} {\bibinfo {author} {\bibfnamefont {G.}~\bibnamefont
  {Giovannetti}}\ and\ \bibinfo {author} {\bibfnamefont {J.}~\bibnamefont
  {van~den Brink}},\ }\href {\doibase 10.1103/PhysRevLett.100.227603}
  {\bibfield  {journal} {\bibinfo  {journal} {Phys. Rev. Lett.}\ }\textbf
  {\bibinfo {volume} {100}},\ \bibinfo {pages} {227603} (\bibinfo {year}
  {2008})}\BibitemShut {NoStop}%
\bibitem [{\citenamefont {Krishnakumar}\ \emph {et~al.}(2001)\citenamefont
  {Krishnakumar}, \citenamefont {Saha-Dasgupta}, \citenamefont {Shanthi},
  \citenamefont {Mahadevan},\ and\ \citenamefont {Sarma}}]{PhysRevB.63.045111}%
  \BibitemOpen
  \bibfield  {author} {\bibinfo {author} {\bibfnamefont {S.~R.}\ \bibnamefont
  {Krishnakumar}}, \bibinfo {author} {\bibfnamefont {T.}~\bibnamefont
  {Saha-Dasgupta}}, \bibinfo {author} {\bibfnamefont {N.}~\bibnamefont
  {Shanthi}}, \bibinfo {author} {\bibfnamefont {P.}~\bibnamefont {Mahadevan}},
  \ and\ \bibinfo {author} {\bibfnamefont {D.~D.}\ \bibnamefont {Sarma}},\
  }\href {\doibase 10.1103/PhysRevB.63.045111} {\bibfield  {journal} {\bibinfo
  {journal} {Phys. Rev. B}\ }\textbf {\bibinfo {volume} {63}},\ \bibinfo
  {pages} {045111} (\bibinfo {year} {2001})}\BibitemShut {NoStop}%
\bibitem [{\citenamefont {Nourafkan}\ and\ \citenamefont
  {Kotliar}(2013)}]{PhysRevB.88.155121}%
  \BibitemOpen
  \bibfield  {author} {\bibinfo {author} {\bibfnamefont {R.}~\bibnamefont
  {Nourafkan}}\ and\ \bibinfo {author} {\bibfnamefont {G.}~\bibnamefont
  {Kotliar}},\ }\href {\doibase 10.1103/PhysRevB.88.155121} {\bibfield
  {journal} {\bibinfo  {journal} {Phys. Rev. B}\ }\textbf {\bibinfo {volume}
  {88}},\ \bibinfo {pages} {155121} (\bibinfo {year} {2013})}\BibitemShut
  {NoStop}%
\bibitem [{\citenamefont {Sakai}\ \emph {et~al.}(2011)\citenamefont {Sakai},
  \citenamefont {Fujioka}, \citenamefont {Fukuda}, \citenamefont {Okuyama},
  \citenamefont {Hashizume}, \citenamefont {Kagawa}, \citenamefont {Nakao},
  \citenamefont {Murakami}, \citenamefont {Arima}, \citenamefont {Baron},
  \citenamefont {Taguchi},\ and\ \citenamefont
  {Tokura}}]{PhysRevLett.107.137601}%
  \BibitemOpen
  \bibfield  {author} {\bibinfo {author} {\bibfnamefont {H.}~\bibnamefont
  {Sakai}}, \bibinfo {author} {\bibfnamefont {J.}~\bibnamefont {Fujioka}},
  \bibinfo {author} {\bibfnamefont {T.}~\bibnamefont {Fukuda}}, \bibinfo
  {author} {\bibfnamefont {D.}~\bibnamefont {Okuyama}}, \bibinfo {author}
  {\bibfnamefont {D.}~\bibnamefont {Hashizume}}, \bibinfo {author}
  {\bibfnamefont {F.}~\bibnamefont {Kagawa}}, \bibinfo {author} {\bibfnamefont
  {H.}~\bibnamefont {Nakao}}, \bibinfo {author} {\bibfnamefont
  {Y.}~\bibnamefont {Murakami}}, \bibinfo {author} {\bibfnamefont
  {T.}~\bibnamefont {Arima}}, \bibinfo {author} {\bibfnamefont {A.~Q.~R.}\
  \bibnamefont {Baron}}, \bibinfo {author} {\bibfnamefont {Y.}~\bibnamefont
  {Taguchi}}, \ and\ \bibinfo {author} {\bibfnamefont {Y.}~\bibnamefont
  {Tokura}},\ }\href {\doibase 10.1103/PhysRevLett.107.137601} {\bibfield
  {journal} {\bibinfo  {journal} {Phys. Rev. Lett.}\ }\textbf {\bibinfo
  {volume} {107}},\ \bibinfo {pages} {137601} (\bibinfo {year}
  {2011})}\BibitemShut {NoStop}%
\bibitem [{\citenamefont {Sakai}\ \emph {et~al.}(2012)\citenamefont {Sakai},
  \citenamefont {Fujioka}, \citenamefont {Fukuda}, \citenamefont {Bahramy},
  \citenamefont {Okuyama}, \citenamefont {Arita}, \citenamefont {Arima},
  \citenamefont {Baron}, \citenamefont {Taguchi},\ and\ \citenamefont
  {Tokura}}]{PhysRevB.86.104407}%
  \BibitemOpen
  \bibfield  {author} {\bibinfo {author} {\bibfnamefont {H.}~\bibnamefont
  {Sakai}}, \bibinfo {author} {\bibfnamefont {J.}~\bibnamefont {Fujioka}},
  \bibinfo {author} {\bibfnamefont {T.}~\bibnamefont {Fukuda}}, \bibinfo
  {author} {\bibfnamefont {M.~S.}\ \bibnamefont {Bahramy}}, \bibinfo {author}
  {\bibfnamefont {D.}~\bibnamefont {Okuyama}}, \bibinfo {author} {\bibfnamefont
  {R.}~\bibnamefont {Arita}}, \bibinfo {author} {\bibfnamefont
  {T.}~\bibnamefont {Arima}}, \bibinfo {author} {\bibfnamefont {A.~Q.~R.}\
  \bibnamefont {Baron}}, \bibinfo {author} {\bibfnamefont {Y.}~\bibnamefont
  {Taguchi}}, \ and\ \bibinfo {author} {\bibfnamefont {Y.}~\bibnamefont
  {Tokura}},\ }\href {\doibase 10.1103/PhysRevB.86.104407} {\bibfield
  {journal} {\bibinfo  {journal} {Phys. Rev. B}\ }\textbf {\bibinfo {volume}
  {86}},\ \bibinfo {pages} {104407} (\bibinfo {year} {2012})}\BibitemShut
  {NoStop}%
\bibitem [{\citenamefont {Giovannetti}\ \emph {et~al.}(2012)\citenamefont
  {Giovannetti}, \citenamefont {Kumar}, \citenamefont {Ortix}, \citenamefont
  {Capone},\ and\ \citenamefont {van~den Brink}}]{PhysRevLett.109.107601}%
  \BibitemOpen
  \bibfield  {author} {\bibinfo {author} {\bibfnamefont {G.}~\bibnamefont
  {Giovannetti}}, \bibinfo {author} {\bibfnamefont {S.}~\bibnamefont {Kumar}},
  \bibinfo {author} {\bibfnamefont {C.}~\bibnamefont {Ortix}}, \bibinfo
  {author} {\bibfnamefont {M.}~\bibnamefont {Capone}}, \ and\ \bibinfo {author}
  {\bibfnamefont {J.}~\bibnamefont {van~den Brink}},\ }\href {\doibase
  10.1103/PhysRevLett.109.107601} {\bibfield  {journal} {\bibinfo  {journal}
  {Phys. Rev. Lett.}\ }\textbf {\bibinfo {volume} {109}},\ \bibinfo {pages}
  {107601} (\bibinfo {year} {2012})}\BibitemShut {NoStop}%
\bibitem [{Note1()}]{Note1}%
  \BibitemOpen
  \bibinfo {note} {It is easier to obtain the distortion vertex, $\partial
  _{\xi }{\protect \bf G}^{-1}$, when the Green function matrix is written in a
  basis which is independent of $\xi $ along the path connecting the
  centro-symmetric system ($\xi =0$) to the real system ($\xi
  =1$).}\BibitemShut {Stop}%
\bibitem [{\citenamefont {Aichhorn}\ \emph {et~al.}(2009)\citenamefont
  {Aichhorn}, \citenamefont {Pourovskii}, \citenamefont {Vildosola},
  \citenamefont {Ferrero}, \citenamefont {Parcollet}, \citenamefont {Miyake},
  \citenamefont {Georges},\ and\ \citenamefont
  {Biermann}}]{PhysRevB.80.085101}%
  \BibitemOpen
  \bibfield  {author} {\bibinfo {author} {\bibfnamefont {M.}~\bibnamefont
  {Aichhorn}}, \bibinfo {author} {\bibfnamefont {L.}~\bibnamefont
  {Pourovskii}}, \bibinfo {author} {\bibfnamefont {V.}~\bibnamefont
  {Vildosola}}, \bibinfo {author} {\bibfnamefont {M.}~\bibnamefont {Ferrero}},
  \bibinfo {author} {\bibfnamefont {O.}~\bibnamefont {Parcollet}}, \bibinfo
  {author} {\bibfnamefont {T.}~\bibnamefont {Miyake}}, \bibinfo {author}
  {\bibfnamefont {A.}~\bibnamefont {Georges}}, \ and\ \bibinfo {author}
  {\bibfnamefont {S.}~\bibnamefont {Biermann}},\ }\href {\doibase
  10.1103/PhysRevB.80.085101} {\bibfield  {journal} {\bibinfo  {journal} {Phys.
  Rev. B}\ }\textbf {\bibinfo {volume} {80}},\ \bibinfo {pages} {085101}
  (\bibinfo {year} {2009})}\BibitemShut {NoStop}%
\bibitem [{Note2()}]{Note2}%
  \BibitemOpen
  \bibinfo {note} {We emphasize that we have an orthonormal projection. In
  either AF or PM phases, the number of orbitals is equal to the number of
  bands within the energy window.}\BibitemShut {Stop}%
\bibitem [{\citenamefont {Wissgott}\ \emph {et~al.}(2012)\citenamefont
  {Wissgott}, \citenamefont {Kune\ifmmode~\check{s}\else \v{s}\fi{}},
  \citenamefont {Toschi},\ and\ \citenamefont {Held}}]{PhysRevB.85.205133}%
  \BibitemOpen
  \bibfield  {author} {\bibinfo {author} {\bibfnamefont {P.}~\bibnamefont
  {Wissgott}}, \bibinfo {author} {\bibfnamefont {J.}~\bibnamefont
  {Kune\ifmmode~\check{s}\else \v{s}\fi{}}}, \bibinfo {author} {\bibfnamefont
  {A.}~\bibnamefont {Toschi}}, \ and\ \bibinfo {author} {\bibfnamefont
  {K.}~\bibnamefont {Held}},\ }\href {\doibase 10.1103/PhysRevB.85.205133}
  {\bibfield  {journal} {\bibinfo  {journal} {Phys. Rev. B}\ }\textbf {\bibinfo
  {volume} {85}},\ \bibinfo {pages} {205133} (\bibinfo {year}
  {2012})}\BibitemShut {NoStop}%
\bibitem [{\citenamefont {Georges}\ \emph {et~al.}(1996)\citenamefont
  {Georges}, \citenamefont {Kotliar}, \citenamefont {Krauth},\ and\
  \citenamefont {Rozenberg}}]{RevModPhys.68.13}%
  \BibitemOpen
  \bibfield  {author} {\bibinfo {author} {\bibfnamefont {A.}~\bibnamefont
  {Georges}}, \bibinfo {author} {\bibfnamefont {G.}~\bibnamefont {Kotliar}},
  \bibinfo {author} {\bibfnamefont {W.}~\bibnamefont {Krauth}}, \ and\ \bibinfo
  {author} {\bibfnamefont {M.~J.}\ \bibnamefont {Rozenberg}},\ }\href {\doibase
  10.1103/RevModPhys.68.13} {\bibfield  {journal} {\bibinfo  {journal} {Rev.
  Mod. Phys.}\ }\textbf {\bibinfo {volume} {68}},\ \bibinfo {pages} {13}
  (\bibinfo {year} {1996})}\BibitemShut {NoStop}%
\bibitem [{\citenamefont {Werner}\ and\ \citenamefont
  {Millis}(2006)}]{PhysRevB.74.155107}%
  \BibitemOpen
  \bibfield  {author} {\bibinfo {author} {\bibfnamefont {P.}~\bibnamefont
  {Werner}}\ and\ \bibinfo {author} {\bibfnamefont {A.~J.}\ \bibnamefont
  {Millis}},\ }\href {\doibase 10.1103/PhysRevB.74.155107} {\bibfield
  {journal} {\bibinfo  {journal} {Phys. Rev. B}\ }\textbf {\bibinfo {volume}
  {74}},\ \bibinfo {pages} {155107} (\bibinfo {year} {2006})}\BibitemShut
  {NoStop}%
\bibitem [{\citenamefont {Haule}(2007)}]{PhysRevB.75.155113}%
  \BibitemOpen
  \bibfield  {author} {\bibinfo {author} {\bibfnamefont {K.}~\bibnamefont
  {Haule}},\ }\href {\doibase 10.1103/PhysRevB.75.155113} {\bibfield  {journal}
  {\bibinfo  {journal} {Phys. Rev. B}\ }\textbf {\bibinfo {volume} {75}},\
  \bibinfo {pages} {155113} (\bibinfo {year} {2007})}\BibitemShut {NoStop}%
\bibitem [{\citenamefont {Haule}\ \emph {et~al.}(2010)\citenamefont {Haule},
  \citenamefont {Yee},\ and\ \citenamefont {Kim}}]{PhysRevB.81.195107}%
  \BibitemOpen
  \bibfield  {author} {\bibinfo {author} {\bibfnamefont {K.}~\bibnamefont
  {Haule}}, \bibinfo {author} {\bibfnamefont {C.-H.}\ \bibnamefont {Yee}}, \
  and\ \bibinfo {author} {\bibfnamefont {K.}~\bibnamefont {Kim}},\ }\href
  {\doibase 10.1103/PhysRevB.81.195107} {\bibfield  {journal} {\bibinfo
  {journal} {Phys. Rev. B}\ }\textbf {\bibinfo {volume} {81}},\ \bibinfo
  {pages} {195107} (\bibinfo {year} {2010})}\BibitemShut {NoStop}%
\bibitem [{Note3()}]{Note3}%
  \BibitemOpen
  \bibinfo {note} {The distortion in the ferroelectric phase lowers the local
  symmetry of the lattice, thus lifting the degeneracies of the $t_{2g}$- and
  $e_g$-orbitals. However, the energy splitting is very small due to the
  smallness of the distortion and we keep using this terminology.}\BibitemShut
  {Stop}%
\bibitem [{Note4()}]{Note4}%
  \BibitemOpen
  \bibinfo {note} {In our LSDA+U calculation, only an effective $U_{eff}=U-J$
  enters the Hamiltonian.}\BibitemShut {Stop}%
\bibitem [{Note5()}]{Note5}%
  \BibitemOpen
  \bibinfo {note} {The magnetic moment in LSDA+U method is larger than in
  LDA+DMFT(AF) because of the more localized atomic-like Wannier basis used in
  LSDA+U, leading to a weaker screening of the interaction.~\cite
  {PhysRevB.81.195107}}\BibitemShut {NoStop}%
\bibitem [{\citenamefont {Ederer}\ \emph {et~al.}(2011)\citenamefont {Ederer},
  \citenamefont {Harris},\ and\ \citenamefont {Kov\'a\ifmmode~\check{c}\else
  \v{c}\fi{}ik}}]{PhysRevB.83.054110}%
  \BibitemOpen
  \bibfield  {author} {\bibinfo {author} {\bibfnamefont {C.}~\bibnamefont
  {Ederer}}, \bibinfo {author} {\bibfnamefont {T.}~\bibnamefont {Harris}}, \
  and\ \bibinfo {author} {\bibfnamefont {R.}~\bibnamefont
  {Kov\'a\ifmmode~\check{c}\else \v{c}\fi{}ik}},\ }\href {\doibase
  10.1103/PhysRevB.83.054110} {\bibfield  {journal} {\bibinfo  {journal} {Phys.
  Rev. B}\ }\textbf {\bibinfo {volume} {83}},\ \bibinfo {pages} {054110}
  (\bibinfo {year} {2011})}\BibitemShut {NoStop}%
\bibitem [{\citenamefont {James M.~Rondinelli}\ \emph
  {et~al.}(2009)\citenamefont {James M.~Rondinelli}, \citenamefont
  {S.~Eidelson},\ and\ \citenamefont {Spaldin}}]{PhysRevB.79.205119}%
  \BibitemOpen
  \bibfield  {author} {\bibinfo {author} {\bibfnamefont {J.}~\bibnamefont
  {James M.~Rondinelli}}, \bibinfo {author} {\bibfnamefont {A.~S.}\
  \bibnamefont {S.~Eidelson}}, \ and\ \bibinfo {author} {\bibfnamefont {N.~A.}\
  \bibnamefont {Spaldin}},\ }\href {\doibase 10.1103/PhysRevB.79.205119}
  {\bibfield  {journal} {\bibinfo  {journal} {Phys. Rev. B}\ }\textbf {\bibinfo
  {volume} {79}},\ \bibinfo {pages} {205119} (\bibinfo {year}
  {2009})}\BibitemShut {NoStop}%
\bibitem [{\citenamefont {Giovannetti}\ and\ \citenamefont
  {Capone}()}]{arXiv:1404.7705}%
  \BibitemOpen
  \bibfield  {author} {\bibinfo {author} {\bibnamefont {Giovannetti}}\ and\
  \bibinfo {author} {\bibfnamefont {M.}~\bibnamefont {Capone}},\ }\href
  {\doibase 10.1103/PhysRevB.90.195113} {\bibinfo  {journal} {Phys. Rev. B}\textbf {\bibinfo
  {volume} {90}},\ \bibinfo {pages} {195113} (\bibinfo {year}
  {2014}) }\BibitemShut {NoStop}%
\bibitem [{\citenamefont {Gonçalves}\ \emph {et~al.}(2014)\citenamefont
  {Gonçalves}, \citenamefont {Correia}, \citenamefont {Lopes}, \citenamefont
  {Araújo},\ and\ \citenamefont {Tavares}}]{JPhysCondensMatter26215401}%
  \BibitemOpen
\bibfield  {journal} {  }\bibfield  {author} {\bibinfo {author} {\bibfnamefont
  {V.~S.}\ \bibnamefont {Gonçalves}, \bibfnamefont {J.~N.~Amaral}}, \bibinfo
  {author} {\bibfnamefont {J.~G.}\ \bibnamefont {Correia}}, \bibinfo {author}
  {\bibfnamefont {A.~M.~L.}\ \bibnamefont {Lopes}}, \bibinfo {author}
  {\bibfnamefont {J.~P.}\ \bibnamefont {Araújo}}, \ and\ \bibinfo {author}
  {\bibfnamefont {P.~B.}\ \bibnamefont {Tavares}},\ }\href {\doibase
  10.1088/0953-8984/26/21/215401} {\bibfield  {journal} {\bibinfo  {journal}
  {Journal of Physics: Condensed Matter}\ }\textbf {\bibinfo {volume} {26}},\
  \bibinfo {pages} {215401} (\bibinfo {year} {2014})}\BibitemShut {NoStop}%
\end{thebibliography}
\end{document}